\documentclass[journal,10pt]{IEEEtran}

\usepackage{amssymb}
\usepackage{amsmath}
\usepackage{cite}
\usepackage{url}
\usepackage{xcolor}
\usepackage{cite,graphicx,amsmath,amssymb}
\usepackage{subfigure}
\usepackage{citesort}
\usepackage{fancyhdr}
\usepackage{mdwmath}
\usepackage{mdwtab}
\usepackage{caption}
\usepackage{amsthm}
\usepackage{amsmath}
\usepackage{bbm}
\usepackage{epstopdf}

\newtheorem{remark}{Remark}
\newtheorem{theorem}{Theorem}

\newtheorem{lemma}{Lemma}

\newtheorem{corollary}{Corollary}

\makeatletter
\def\ScaleIfNeeded{%
\ifdim\Gin@nat@width>\linewidth \linewidth \else \Gin@nat@width
\fi } \makeatother

\begin{document}

\title{Semi-Grant-Free NOMA: Ergodic Rates Analysis with Random Deployed Users}


\author{
Chao~Zhang,~\IEEEmembership{Student Member,~IEEE},
Yuanwei~Liu,~\IEEEmembership{Senior Member,~IEEE},
Wenqiang~Yi,~\IEEEmembership{Student Member,~IEEE},
Zhijin~Qin,~\IEEEmembership{Member,~IEEE} and
Zhiguo~Ding,~\IEEEmembership{Fellow,~IEEE}

\vspace{-0.5cm}
\thanks{C. Zhang, Y. Liu, W. Yi and Z. Qin are with the School of Electronic Engineering and Computer Science, Queen Mary University of London, London, UK (email:\{chao.zhang, yuanwei.liu, w.yi, z.qin\}@qmul.ac.uk).}
\thanks{Z. Ding is with the School of Electrical and Electronic Engineering, The University of Manchester, Manchester, UK (e-mail: zhiguo.ding@manchester.ac.uk).}

}

\maketitle

\begin{abstract}
Semi-grant-free (Semi-GF) non-orthogonal multiple access (NOMA) enables grant-free (GF) and grant-based (GB) users to share the same resource blocks, thereby balancing the connectivity and stability of communications. This letter analyzes ergodic rates of Semi-GF NOMA systems. First, this paper exploits a Semi-GF protocol, denoted as dynamic protocol, for selecting GF users into the occupied GB channels via the GB user's instantaneous received received power. Under this protocol, the closed-form analytical and approximated expressions for ergodic rates are derived. The numerical results illustrate that the GF user (weak NOMA user) has a performance upper limit, while the ergodic rate of the GB user (strong NOMA user) increases linearly versus the transmit signal-to-noise ratio.
\end{abstract}

\begin{IEEEkeywords}
{D}ynamic protocol, ergodic rate, semi-grant-free, uplink NOMA
\end{IEEEkeywords}
\vspace{-0.4cm}
\section{Introduction}

Grant-free (GF) is a promising technique for scenarios in the fifth generation (5G) and beyond, especially for ultra-reliable low latency communications (URLLC). With cancelations of uplink scheduling requests (SR) and dynamic scheduling grants (SG) from conventional grant-based (GB) transmission, GF enhances latency of transmission and utility efficiency of resource blocks. In spite of the aforementioned advantages of GF transmission, GB transmission is still irreplaceable, especially for primary users requiring assured transmission. Therefore, we need to consider the practical scenario that the GB and GF schemes simultaneously exist. The traditional solution is to allocate different spectrum resources to GB and GF transmission. However, several challenges emerge: 1) the reserved spectrum resources are not affordable for explosively increasing GB users; 2) channels of GF users are occasionally occupied with low utility efficiency since GF users always transmit small packets within a limited transmitting period. As a consequence, it is essential to figure out an efficient spectrum allocation method. To this end, it can be allowed that the GF and GB users to share the same spectrum resources. After that, the sparse channels of GF users can be efficiently utilized by the GB users whose available spectrum resources are enlarged. This new technique is named as Semi-grant-free (Semi-GF) transmission \cite{st32,semigf2}.

As ubiquitous connectivity can be anticipated as the trend in future networks, collisions under Semi-GF transmission is still frequently incurred in multi-user scenarios \cite{GFGB}. To resolve this issue, non-orthogonal multiple access (NOMA) technique is exploited to achieve the mitigation of collisions \cite{yuanwei5GNOMA}. With multiplexing strategies and successive interference cancellation (SIC) techniques, multiple devices can be connected successfully without collision in the same resource block \cite{8635489}. For Semi-GF NOMA, \cite{st32} has proposed two protocols, i.e., open-loop protocol and distributed protocol, which employ average channel gain as thresholds to select GF users joining into the channels occupied by GB users. This protocol focuses on long-term communications with long data streams.
When considering short data packets that can be transmitted within a coherent time, the instantaneous channel gain is dominant for NOMA user pairing since the channel gains for different transmission are various \cite{ICSI2}. Hence, this work utilizes the GB user's instantaneous received power as accurate thresholds. The Semi-GF protocol is defined as the dynamic protocol. Therefore, it is considered that Semi-GF NOMA systems under the dynamic protocol to achieve low latency, enhanced spectrum efficiency, and increased connectivity.

This paper investigates ergodic rates of the proposed Semi-GF NOMA systems. The contributions are summarized as follows: 1) {this paper considers the GB user's instantaneous received power to select GF users into NOMA pairs via the dynamic protocol;} 2) the closed-form expressions of ergodic rates for GF and GB users are derived; 3) it can be observed that the ergodic rate of the GB users has no error floors and the ergodic rate of the GF users has a maximum value.

\vspace{-0.35cm}

\section{System model}

With a fixed base station (BS) at the center of the disc, the GF and GB users are uniformly distributed in the disc with the radium as $R_L$. The distances from the users to the BS is denoted as $d_{G}$, where $G \in\{GF,GB\}$ means various users. It is assumed that the GB users have been allocated into the dedicated orthogonal resource blocks, thereby the CSI of the GB users has been obtained and the interference from other pairs is omitted \cite{TWC}. To maintain the performance of SIC process, only one selected GF user and one connected GB user are randomly paired into the same orthogonal resource block.  Under Semi-GF protocols, the GB and GF users can share the same resource block via NOMA technique.

\vspace{-0.4cm}
\subsection{Semi-GF Protocol}

This paper exploits a Semi-GF protocol, denoted as dynamic protocol. Firstly, several assumptions for the dynamic protocol are clarified as follows:

\vspace{-0.1cm}
{\begin{itemize}
 \item We considers the connected GB user is a delay-sensitive user transmitting messages such as voice calls or emergency healthcare status.
 \item We consider GF users transmit short packets in limited transmitting time, which is less than the coherent time. Thus, the channel gains of GB users are evaluated as a constant. Hence, the instantaneous received power of the connected GB users is exploited as thresholds to select GF users \cite{st32}.
 \item A GF user with lower received power at their BS compared to the connected GB user is chosen into NOMA pairs since the delay-sensitive GB user needs to avoid the latency period caused by the SIC procedure.
\end{itemize}}

{The procedure of dynamic protocol\footnote{{It is noted the users' ergodic rates are of interest. If the GB user has a fixed data rate, the constraint shown in \eqref{constrait} cannot guarantee that the GB user's target data rate can be met, which is different from the existing one in \cite{st32}.}} is summarized that: 1) the BS broadcasts its transmit power and the instantaneous received signal power of GB users as a threshold to GF users; 2) the GF users first use the transmit power of the BS and the received signal power from the BS to estimate their received signal power at the BS end (RPaB). Then, they complete a comparison between their RPaB and the threshold; 3) the GF user with lower RPaB will access into the resource block of the connected GB user to form a NOMA pair. The constraint\footnote{{With $n$ GF accessing in NOMA channels, we assume the sum load of GF users in each resource block does not exceed the max load, which controls the interference from GF users to the GB user to maintain the SIC process. Under this condition, the constraint is expressed as ${P_{GF}}{\left| {{h_{G{F_{n}}}}} \right|^2}d_{G{F_{n}}}^{ - \alpha }$ $ <  \cdots  < {P_{GF}}{\left| {{h_{G{F_1}}}} \right|^2}d_{G{F_1}}^{ - \alpha } < {P_{GB}}{\left| {{h_{GB}}} \right|^2}d_{GB}^{ - \alpha }$.}} can be mathematically expressed as
\begin{align}\label{constrait}
\underbrace {{P_{GF}}{{\left| {{h_{GF}}} \right|}^2}d_{GF}^{ - \alpha }}_{{\rm{RPaB}}} < \underbrace {{P_{GB}}{{\left| {{h_{GB}}} \right|}^2}d_{GB}^{ - \alpha }}_{{\rm{threshold}}},
\end{align}
where $P_G$ is the transmit power of users, $\alpha$ is the path loss exponent and $\left |h_G\right|^2$ with $G \in \{GF,GB\}$ is the Rayleigh fading channels with mean values $\lambda_{GF}=1$ and $\lambda_{GB}=1$.}

\vspace{-0.2cm}
\subsection{SINR Analysis}

Under dynamic protocol, the signal-to-interference-plus-noise ratio (SINR) of the $GF$ and $GB$ user is expressed as
\begin{align}\label{gamma_GB}
&\gamma _{GB}^{} = \frac{{{P_{GB}}{{\left| {{h_{GB}}} \right|}^2}d_{GB}^{ - \alpha }}}{{{P_{GF}}{{\left| {{h_{GF}}} \right|}^2}d_{GF}^{ - \alpha } + {\sigma ^2}}},\gamma _{GF}^{} = \frac{{{P_{GF}}{{\left| {{h_{GF}}} \right|}^2}d_{GF}^{ - \alpha }}}{{{\sigma ^2}}},
\end{align}
where $\sigma^{2}$ is the variance of additive white Gaussian noise.

\vspace{-0.3cm}
\section{Ergodic Rates }

{Ergodic rates indicate the averaged achievable rate, with the definition of ${\rm{E}}\left[ {{R_G}} \right] = {\rm{E}}\left[ {{{\log }_2}\left( {1 + {\gamma _G}} \right)} \right]$, where $\rm{E}[\cdot]$ is calculating the expectation with $G \in\{GF,GB\}$. Thus, the high ergodic rate achieves superior performance, i.e., high channel capacity. Under dynamic protocol, the ergodic rates of the GB and GF users are expressed as
\begin{align}\label{ERndef}
&{\rm{E}}\left[ {{R_{GB}}} \right] = \int_0^\infty  {\frac{{P_c^{GB}\left( \gamma _{GB}^{th} \right)}}{{1 + \gamma _{GB}^{th}}}d\gamma _{GB}^{th}} ,\\
\label{ERmdef}
&{\rm{E}}\left[ {{R_{GF}}} \right]=\int_0^\infty  {\frac{{P_c^{GF}\left( \gamma _{GF}^{th} \right)}}{{1 + \gamma _{GF}^{th}}}d\gamma _{GF}^{th}} ,
\end{align}
where $\gamma _{GB}^{th}$ and $\gamma _{GF}^{th}$ are the outage thresholds of users, ${g_{GF}} = {P_{GF}}{\left| {{h_{GF}}} \right|^2}d_{GF}^{ - \alpha }$, ${g_{GB}} = {P_{GB}}{\left| {{h_{GB}}} \right|^2}d_{GB}^{ - \alpha }$, $P_c^{GB}$ and $P_c^{GF}$ are the coverage probability of users, which are expressed as
 \begin{align}
&P_c^{GB} = \Pr \left\{ {\gamma _{GB}^{} > \gamma _{GB}^{th},{g_{GF}} < {g_{GB}}} \right\},\\
&P_c^{GF} = \Pr \left\{ {\gamma _{GB}^{} > \gamma _{GB}^{th},\gamma _{GF}^{} > \gamma _{GF}^{th},{g_{GF}} < {g_{GB}}} \right\}.
 \end{align}}

\vspace{-0.3cm}
\subsection{Exact Analysis on Ergodic Rates}
\vspace{-0.1cm}
In this subsection, the exact closed-form expressions of ergodic rates are derived for the GF and GB users respectively, shown as \textbf{Theorem \ref{ERGB}} and \textbf{Theorem \ref{ERGF}}.

\subsubsection{Ergodic Rates for the GB user}
Before deriving the final expressions of ergodic rates for the GB users, two scenarios are analyzed as \textbf{Lemma \ref{lemma1}} with $\gamma_{GB}^{th} \in [1,\infty]$ and \textbf{Lemma \ref{lemma2}} with $\gamma_{GB}^{th} \in [0,1]$ since the derivations are different. For the first case, signals can be decoded as the signal strength is stronger than the interference strength. For the second case, although the signal strength is weaker than that of interference, error correction coding can be used to help the decoding.
\vspace{-0.1cm}
\begin{lemma}\label{lemma1}
\emph{Conditioned on $\gamma_{GB}^{th} \in [1,\infty]$, the closed-form expressions of ergodic rates for the GB users are derived as }
\begin{align}
{\rm{E}}\left[ {{R_{GB}}} \right]_{1} &= 2R_L^{\alpha  - 1}\sum\limits_{n = 1}^N {\sum\limits_{m = 1}^M {\frac{{{\omega _n}\sqrt {1 + \varepsilon _n^2} {\omega _m}\sqrt {1 + \varepsilon _m^2} }}{{{{\left( {{\varepsilon _m} + 1} \right)}^2}\left( {1 + \frac{1}{\alpha }} \right){\Theta _1}}}} } \notag\\
 &\hspace*{0.3cm}\times \frac{{\exp \left( { - {\Theta _2}\Omega _1^\alpha \left( {{\varepsilon _n}} \right){\Omega _2}\left( {{\varepsilon _m}} \right)} \right)}}{{\Omega _1^{\alpha  - 1}\left( {{\varepsilon _n}} \right)\left( {{\Omega _2}\left( {{\varepsilon _m}} \right) + \Omega _2^2\left( {{\varepsilon _m}} \right)} \right)}}\notag\\
 &\hspace*{0.3cm}\times {}_2{F_1}\left( {1,\frac{{2 + \alpha }}{\alpha };2 + \frac{2}{\alpha }; - \frac{{R_L^\alpha \Omega _1^{ - \alpha }\left( {{\varepsilon _n}} \right)}}{{{\Theta _1}{\Omega _2}\left( {{\varepsilon _m}} \right)}}} \right),
\end{align}
\emph{where ${}_2{F_1}(\cdot)$ is the hypergeometric function in \cite{Prudnikov}, ${\Omega _1}\left( x \right) = \frac{{{R_L}}}{2}\left( {x + 1} \right)$, ${\Omega _2}\left( x \right) = \frac{2}{{x + 1}}$, ${\varepsilon _n} = \cos \left( {\frac{{2n - 1}}{{2N}}\pi } \right)$, ${\varepsilon _m} = \cos \left( {\frac{{2m - 1}}{{2M}}\pi } \right)$, ${\omega _n} = \pi /N$, ${\omega _i} = \pi /M$, $M$ and $N$ is the coefficients in Chebyshev-Gauss quadrature, ${\Theta _1} = \frac{{{\lambda _{GF}}{P_{GF}}}}{{{\lambda _{GB}}{P_{GB}}}}$ and ${\Theta _2} = \frac{{{\sigma ^2}}}{{{\lambda _{GB}}{P_{GB}}}}$. }
\end{lemma}
\begin{IEEEproof}
See Appendix~A.
\end{IEEEproof}
\vspace{-0.1cm}

\begin{lemma}\label{lemma2}
\emph{{Note that messages can be decoded via error correction coding with redundant information when $\gamma_{GB}^{th} \in [0,1]$ for the GB user \cite{Zhiguo}. Thus, the closed-form expressions of ergodic rates can be derived as }}
\begin{align}
{\rm{E}}\left[ {{R_{GB}}} \right]_{2} = {{\rm I}_3}+{{\rm I}_4},
\end{align}
\emph{where ${{\rm I}_3}$ and ${{\rm I}_4}$ is the first and second items in \eqref{B1} as }
\begin{align}
{{\rm I}_3} &= \sum\limits_{n = 1}^N {\sum\limits_{m = 1}^M {\frac{{{\omega _n}\sqrt {1 + \varepsilon _n^2} {\omega _m}\sqrt {1 + \varepsilon _m^2} }}{{\ln 2{\lambda _{GF}}R_L^2}}} } {\Omega _1}\left( {{\varepsilon _n}} \right){\Omega _1}\left( {{\varepsilon _m}} \right)\notag\\
&\hspace*{0.3cm}\times{\delta _1}\left( {{\Omega _1}\left( {{\varepsilon _n}} \right),{\Omega _1}\left( {{\varepsilon _m}} \right)} \right)\Xi \left( {{\varepsilon _n},{\varepsilon _m}} \right),\\
{{\rm I}_4} &= \sum\limits_{n = 1}^N {\sum\limits_{m = 1}^M {\frac{{{\omega _n}\sqrt {1 + \varepsilon _n^2} {\omega _m}\sqrt {1 + \varepsilon _m^2} {\Omega _1}\left( {{\varepsilon _n}} \right){\Omega _1}\left( {{\varepsilon _m}} \right)}}{{R_L^{\rm{2}}{\lambda _{GF}}\Psi \left( {{\Omega _1}\left( {{\varepsilon _n}} \right),{\Omega _1}\left( {{\varepsilon _m}} \right),{\rm{1}}} \right)}}} } \notag\\
 &\hspace*{0.3cm}\times \exp \left( { - \Psi \left( {{\Omega _1}\left( {{\varepsilon _n}} \right),{\Omega _1}\left( {{\varepsilon _m}} \right),{\rm{1}}} \right)\frac{{{\sigma ^2}\Omega _{\rm{1}}^\beta \left( {{\varepsilon _n}} \right)}}{{{P_{GF}}}}} \right),
\end{align}
\emph{where $\Psi \left( {y,z,t} \right) = \left( {\frac{{{P_{GF}}{y^{ - \alpha }}t}}{{{\lambda _{GB}}{P_{GB}}{z^{ - \alpha }}}} + \frac{1}{{{\lambda _{GF}}}}} \right)$, $\Xi \left( {{\varepsilon _n},{\varepsilon _m}} \right) = \Phi \left( {1,{\delta _2}\left( {{\Omega _1}\left( {{\varepsilon _m}} \right)} \right)} \right) - {\delta _3}\left( {{\Omega _1}\left( {{\varepsilon _n}} \right)} \right)\Phi \left( {1,2{\delta _2}\left( {{\Omega _1}\left( {{\varepsilon _m}} \right)} \right)} \right) - \Phi \left( {{\delta _4}\left( {{\Omega _1}\left( {{\varepsilon _n}} \right),{\Omega _1}\left( {{\varepsilon _m}} \right)} \right),{\delta _2}\left( {{\Omega _1}\left( {{\varepsilon _m}} \right)} \right)} \right) + {\delta _3}\left( {{\Omega _1}\left( {{\varepsilon _n}} \right)} \right)\times$ $\Phi \left( {{\delta _4}\left( {{\Omega _1}\left( {{\varepsilon _n}} \right),{\Omega _1}\left( {{\varepsilon _m}} \right)} \right),2{\delta _2}\left( {{\Omega _1}\left( {{\varepsilon _m}} \right)} \right)} \right)$, ${\delta _1}\left( {y,z} \right) = \frac{{{\lambda _{GF}}{\lambda _{GB}}{P_{GB}}{z^{ - \alpha }}}}{{\left( {{\lambda _{GB}}{P_{GB}}{z^{ - \alpha }} - {\lambda _{GF}}{P_{GF}}{y^{ - \alpha }}} \right)}}$, ${\delta _2}\left( z \right) = \frac{{{\sigma ^2}}}{{{\lambda _{GB}}{P_{GB}}{z^{ - \alpha }}}}$, ${\delta _3}\left( y \right) = \exp \left( { - \frac{{{\sigma ^2}{y^\alpha }}}{{{\lambda _{GF}}{P_{GF}}}}} \right)$, ${\delta _4}\left( {y,z} \right) = \frac{{{\lambda _{GB}}{P_{GB}}{z^{ - \alpha }}}}{{{\lambda _{GF}}{P_{GF}}{y^{ - \alpha }}}}$, $\Phi \left( {a,b} \right) = \int_0^1 {\frac{{\exp \left( { - bt} \right)}}{{t + a}}} dt =  - \exp \left( {ab} \right){\rm{Ei}}\left( { - ab} \right)$ and ${\rm{Ei}}\left( x \right) = \int_{ - \infty }^x {\frac{{\exp (t)}}{t}} dt$ is the exponential integral in \cite{Prudnikov}. Other coefficients are defined in \textbf{Lemma \ref{lemma1}}.  }
\end{lemma}
\begin{IEEEproof}
See Appendix~B.
\end{IEEEproof}

\begin{theorem}\label{ERGB}
\emph{Combined the expressions of ergodic rate in two scenarios shown in \textbf{Lemma \ref{lemma1}} and \textbf{Lemma \ref{lemma2}}, the final ergodic rate expression for the GB user can be expressed as}
\begin{align}
{\rm{E}}\left[ {{R_{GB}}} \right]={\rm{E}}\left[ {{R_{GB}}} \right]_{1}+{\rm{E}}\left[ {{R_{GB}}} \right]_{2}.
\end{align}
\end{theorem}

\begin{remark}\label{R1}
{We observe that the ergodic rate of the GB user is proportional to the transmit power of the GB user $P_{GB}$ but inversely proportional to the transmit power of GF users $P_{GF}$.}
\end{remark}

\subsubsection{Ergodic Rates for the GF user}
the GF user should satisfy two conditions that: 1) received power limit, i.e., ${P_{GF}}{\left| {{h_{GF}}} \right|^2}d_{GF}^{ - \alpha } < {P_{GB}}{\left| {{h_{GB}}} \right|^2}d_{GB}^{ - \alpha }$ and, 2) a success SIC procedure, i.e., $\gamma _{GB}^{} > \gamma _{GB}^{th}$. Based on those two conditions, the closed-form ergodic rate expressions for the GF user are derived.

\begin{theorem}\label{ERGF}
\emph{Under the two conditions mentioned above, the closed-form expressions of ergodic rates for the GF user can be derived as }
\begin{align}
&E\left[ {R_{GF}^{}} \right] = \sum\limits_{n = 1}^N {\sum\limits_{m = 1}^M {\frac{{ - {\omega _n}\sqrt {1 + \varepsilon _n^2} {\omega _m}\sqrt {1 + \varepsilon _m^2} }}{{\ln 2{\lambda _{GF}}R_L^2}}} }\notag \\
& \times \frac{{{\Omega _1}\left( {\varepsilon _n^{}} \right){\Omega _1}\left( {\varepsilon _m^{}} \right)\exp \left( { - \frac{{{\sigma ^2}\gamma _{GB}^{th}\Omega _1^\alpha \left( {\varepsilon _m^{}} \right)}}{{{\lambda _{GB}}{P_{GB}}}}} \right)}}{{\Psi \left( {{\Omega _1}\left( {\varepsilon _n^{}} \right),{\Omega _1}\left( {\varepsilon _m^{}} \right),\gamma _{GB}^{th}} \right)}}\notag\\
& \times \exp \left( {\Psi \left( {{\Omega _1}\left( {\varepsilon _n^{}} \right),{\Omega _1}\left( {\varepsilon _m^{}} \right),\gamma _{GB}^{th}} \right)\frac{{{\sigma ^2}\Omega _1^\alpha \left( {\varepsilon _n^{}} \right)}}{{{P_{GF}}}}} \right)\notag\\
& \times {\rm{Ei}}\left( { - \Psi \left( {{\Omega _1}\left( {\varepsilon _n^{}} \right),{\Omega _1}\left( {\varepsilon _m^{}} \right),\gamma _{GB}^{th}} \right)\frac{{{\sigma ^2}\Omega _1^\alpha \left( {\varepsilon _n^{}} \right)}}{{{P_{GF}}}}} \right),
\end{align}
\emph{where coefficients are defined in \textbf{Lemma \ref{lemma1}} and \textbf{Lemma \ref{lemma2}}.}
\end{theorem}
\begin{IEEEproof}
See Appendix~C.
\end{IEEEproof}

\begin{remark}\label{R2}
{Analytical results indicate the ergodic rate of the GF user is inversely proportional to $P_{GB}$. When $P_{GF}$ enhances, although the SNR of GF users will increase, the probability of GF users joining into the occupied GB channel is reduced, thereby a maximum ergodic rate of GF users is obtained.}
\end{remark}

\vspace{-0.25cm}
\subsection{Approximated Analysis on Ergodic Rates}
This paper investigates the approximated ergodic rate for GF users via \textbf{Corollary \ref{corollary1}} in this subsection.

\begin{figure}[!htb]
\vspace{-0.3cm}
\centering
\includegraphics[width= 3in]{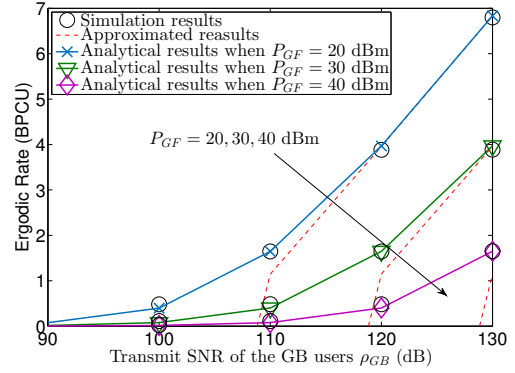}
\caption{{Ergodic rates (BPCU) of the GB user v.s. transmit SNR of GB users $\rho_{GB}$ (dB)}}
\vspace{-0.3cm}
\label{Fig1}
\end{figure}
\begin{figure}[!htb]
\vspace{-0.1cm}
\centering
\includegraphics[width= 3in]{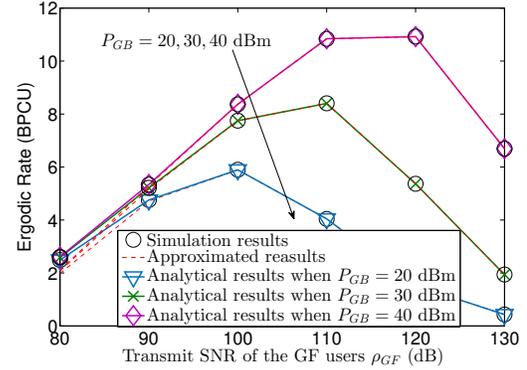}
\caption{{Ergodic rates (BPCU) of the GF user v.s. transmit SNR of GF user $\rho_{GF}$ (dB)} }
\vspace{-0.5cm}
\label{Fig2}
\end{figure}

\begin{corollary}\label{corollary1}
\emph{It is assumed that the transmit power of the GF user to infinity, denoted as $\rho_{GF}=P_{GF}/\sigma^2 \to \infty$. With the aid of asymptotic expressions, the approximated expression of the ergodic rate for the GF user can be calculated as  }
\begin{align}
&E\left[ {R_{GF}^{}} \right] = \sum\limits_{n = 1}^N {\sum\limits_{m = 1}^M {\frac{{ - {\omega _n}\sqrt {1 + \varepsilon _n^2} {\omega _m}\sqrt {1 + \varepsilon _m^2} }}{{\ln 2{\lambda _{GF}}R_L^2}}} }\notag \\
& \times \frac{{{\Omega _1}\left( {\varepsilon _n^{}} \right){\Omega _1}\left( {\varepsilon _m^{}} \right)\exp \left( { - \frac{{{\sigma ^2}\gamma _{GB}^{th}\Omega _1^\alpha \left( {\varepsilon _m^{}} \right)}}{{{\lambda _{GB}}{P_{GB}}}}} \right)}}{{\Psi \left( {{\Omega _1}\left( {\varepsilon _n^{}} \right),{\Omega _1}\left( {\varepsilon _m^{}} \right),\gamma _{GB}^{th}} \right)}}\notag\\
& \times \left( {\ln \left( {\Psi \left( {{\Omega _1}\left( {\varepsilon _n^{}} \right),{\Omega _1}\left( {\varepsilon _m^{}} \right),\gamma _{GB}^{th}} \right)\frac{{{\sigma ^2}\Omega _1^\alpha \left( {\varepsilon _n^{}} \right)}}{{{P_{GF}}}}} \right) + C} \right)\notag\\
& \times \left( {{\rm{1}} - \Psi \left( {{\Omega _1}\left( {\varepsilon _n^{}} \right),{\Omega _1}\left( {\varepsilon _m^{}} \right),\gamma _{GB}^{th}} \right)\frac{{{\sigma ^2}\Omega _1^\alpha \left( {\varepsilon _n^{}} \right)}}{{{P_{GF}}}}} \right).
\end{align}
\end{corollary}
\begin{IEEEproof}
When assumed $x \to 0$, asymptotic expressions such as $1-e^{-x}=x$ and $\rm{Ei}(-x)=ln(x)+C$ can be utilized, where $C \approx 0.577215$ as a constant in \cite{Prudnikov}.
\end{IEEEproof}

Due to space limitations, the approximated ergodic rate expressions of the GB users are omitted. It is presented that the simulation results of approximated expressions via dashed lines in the next section.

\vspace{-0.3cm}
\section{Numerical Results}

Numerical results are indicated to validate analytical ergodic rates (\textbf{Theorem \ref{ERGB}} and \textbf{Theorem \ref{ERGF}}) with the unit as bit per channel use (BPCU), which are demonstrated as close agreements. Without otherwise specification, the numerical settings are defined as: the rudium of the disc as $600$ m, the calculated noise power as ${\sigma ^2} =  - 170 + 10\log \left( {BW} \right) + {N_f}=-90 $ dB, where $BW$ is the bandwidth as $10$ MHz and $N_f$ is noise figure as $10$ dB, path loss exponent as $2.8$.

In Fig. 1, it depicts the ergodic rates of the GB users versus $\rho_{GB} = P_{GB}/\sigma^2 \in [0,40]$ dB with various transmit power of the GF user $P_{GF} = \{20,40\}$ dBm. The dashed lines are the approximated ergodic rates, which match simulation results in high SNR regions. One observation is that the performance of the GB users has no error floors. Thus, allocating combinations of transmit powers for the GF and GB users can avoid performance deterioration. Another observation is that the ergodic rates are reduced when the transmit power $P_{GF}$ increases since the GF users add interference to the GB user.

In Fig. 2, this paper investigates the ergodic rate performance corresponding to the GF users versus $\rho_{GF} = P_{GF}/\sigma^2 \in [0,40]$ with transmit power of the GB user $P_{GB} = \{20,40\}$ dBm. The dashed lines represent approximated ergodic rate expressions. It is observed that the ergodic rate performance has the maximum value. After reaching the maximum value, the ergodic rates decrease when enhancing the GF users' transmit power. This is because when the transmit powers of GF users increase, the SIC procedure will have high outage situations, which cause the message of the GF users cannot be decoded.

\vspace{-0.3cm}
\section{Conclusions}
This paper has investigated ergodic rates in Semi-GF NOMA systems. {The dynamic protocol has been exploited by utilizing instantaneous received power as thresholds.} The closed-form expressions of the ergodic rates for the GB and GF users have been derived. Numerical results reveal that 1) the performance of the GB users can be maintained since there is no error floor for the ergodic rates of the GB users; 2) the performance of the GF users has an upper limit, which depends on the performance of SIC procedure.

\numberwithin{equation}{section}

\section*{Appendix~A: proof of Lemma \ref{lemma1}} \label{Appendix:A}
\renewcommand{\theequation}{A.\arabic{equation}}
\setcounter{equation}{0}

{When $\gamma_{GB}^{th} \in [1,\infty]$, the ergodic rate expression can be rewritten via integrations as
\begin{align}\label{A1}
&E\left[ {R_{GB}^{}} \right] = \frac{1}{{\ln 2}}\int_1^\infty  {\frac{1}{{1 + t}}} \int_0^{{R_L}} {\int_0^{{R_L}} {} }\notag \\
&\underbrace { \times \int_0^\infty  {\left( {1 - {F_{{{\left| {{h_{GB}}} \right|}^2}}}\left( {\frac{{{P_{GF}}tx{y^{ - \alpha }} + {\sigma ^2}t}}{{{P_{GB}}{z^{ - \alpha }}}}} \right)} \right)} {f_{{{\left| {{h_{GF}}} \right|}^2}}}\left( x \right)dx}_{{{\rm I}_1}}\notag\\
& \times {f_{d_{GF}}}\left( y \right)dy{f_{d_{GB}}}\left( z \right)dzdt,
\end{align}
where all the channels obey Rayleigh fading channels and $ f_{d_{GF}}(x) = f_{d_{GB}}(x) = 2x/\left(R_L^2\right)$.}

With the aid of Eq. [3.194.1] in \cite{Prudnikov} and hypergeometric functions, $\rm{I_2}$ can be expressed and derived as
\begin{align} \label{I2}
 {{\rm I}_2}&= \int_0^{{R_L}} {\frac{{2y{{\rm I}_1}}}{{R_L^2}}}dy = \frac{{R_L^\alpha \exp \left( { - {\Theta _2}t{z^\alpha }} \right)}}{{\left( {1 + \frac{1}{\alpha }} \right){\Theta _1}t{z^\alpha }\left( {1 + t} \right)}}\notag\\
& \times {}_2{F_1}\left( {1,\frac{{2 + \alpha }}{\alpha };2 + \frac{2}{\alpha }; - \frac{{R_L^\alpha }}{{{\Theta _1}t{z^\alpha }}}} \right).
\end{align}

The Chebyshev-Gauss-quadrature can be defined as $\int_{ - 1}^1 {\frac{{f\left( x \right)}}{{\sqrt {1 - {x^2}} }}dx}  \approx \sum\nolimits_{i = 1}^N {{\omega _i}f\left( {{x_i}} \right)} $, where ${x_i} = \cos \left( {\frac{{2i - 1}}{{2I}}\pi } \right)$ and ${\omega _i} = \pi /I$. Substituting $\rm{I_2}$ into \eqref{A1} and harnessing Chebyshev-Gauss-quadrature, the final expressions of ergodic rate are obtained for the GB user with $\gamma_{GB}^{th} \in[1,\infty]$.

\vspace{-0.3cm}
\section*{Appendix~B: Proof of Lemma~\ref{lemma2}} \label{Appendix:B}
\renewcommand{\theequation}{B.\arabic{equation}}
\setcounter{equation}{0}

When $\gamma _{GB}^{th} < 1$, the integration expressions of ergodic rates can be expressed as
\begin{align}\label{B1}
&E{\left[ {R_{GB}^{}} \right]_2} = \frac{1}{{\ln 2}}\int_0^1 {\int_0^{{R_L}} {\int_0^{{R_L}} {\frac{{\exp \left( { - \frac{{t{\sigma ^2}}}{{{\lambda _{GB}}{P_{GB}}{z^{ - \alpha }}}}} \right)}}{{1 + t}}} } } \notag\\
& \times \int_0^{\frac{{{\sigma ^2}{y^\alpha }}}{{{P_{GF}}}}} {\exp \left( { - \Psi \left( {y,z,t} \right)x} \right)dx} {f_{d_{GF}^{ - \alpha }}}\left( y \right)dy{f_{d_{GB}^{ - \alpha }}}\left( z \right)dzdt\notag\\
 &+ \int_0^{{R_L}} {\int_0^{{R_L}} {\int_{\frac{{{\sigma ^2}{y^\alpha }}}{{{P_{GF}}}}}^\infty  {\exp \left( { - \frac{{{P_{GF}}x{y^{ - \alpha }}}}{{{\lambda _{GB}}{P_{GB}}{z^{ - \alpha }}}}} \right){f_{{{\left| {{h_{GF}}} \right|}^2}}}\left( x \right)dx} } } \notag\\
 &\times {f_{d_{GF}^{ - \alpha }}}\left( y \right)dy{f_{d_{GB}^{ - \alpha }}}\left( z \right)dz.
\end{align}

Denoted the first item of \eqref{B1} as $\rm{I_3}$. For the first item, based on Eq. [1.3.2.22] and Eq. [2.3.4.3] in \cite{Prudnikov}, the ergodic rate expressions can be derived as
\begin{align}\label{B2}
 \rm{I_3}&= \frac{4}{{\ln 2{\lambda _{GF}}R_L^4}}\int_0^{{R_L}} {\int_0^{{R_L}} {yz{\delta _1}\left( {y,z} \right)\left\{ {\Phi \left( {1,{\delta _2}\left( z \right)} \right)} \right.} } \notag\\
 &\hspace*{0.3cm}- {\delta _3}\left( y \right)\Phi \left( {1,2{\delta _2}\left( z \right)} \right) - \Phi \left( {{\delta _4}\left( {y,z} \right),{\delta _2}\left( z \right)} \right)\notag\\
& \hspace*{0.3cm}+ \left. {{\delta _3}\left( y \right)\Phi \left( {{\delta _4}\left( {y,z} \right),2{\delta _2}\left( z \right)} \right)} \right\}dydz.
\end{align}

We denote the second item of \eqref{B1} as $\rm{I_4}$. The equation $\rm{I_4}$ is derived as
\begin{align}\label{B3}
{{\rm I}_4} = \frac{4}{{R_L^4}}\int_0^{{R_L}} {\int_0^{{R_L}} {\frac{{yz\exp \left( { - \Psi \left( {y,z,t} \right)\frac{{{\sigma ^2}{y^\alpha }}}{{{P_{GF}}}}} \right)}}{{{\lambda _{GF}}\Psi \left( {y,z,t} \right)}}} } dydz
\end{align}

Substituting \eqref{B2} and \eqref{B3} into \eqref{B1} and utilizing Chebyshev-Gauss quadrature, the final closed-form expressions can be achieved.

\section*{Appendix~C: Proof of Theorem~\ref{ERGF}} \label{Appendix:C}
\renewcommand{\theequation}{C.\arabic{equation}}
\setcounter{equation}{0}

{Under the conditions of SIC procedure and dynamic protocol, the expressions of ergodic rate for the GF user can be expressed as
\begin{align}\label{C1}
&E\left[ {R_{GF}^{}} \right] = \int_0^\infty  {\int_0^{{R_L}} {\int_0^{{R_L}} {\int_{\frac{{{\sigma ^2}t{y^\alpha }}}{{{P_{GF}}}}}^\infty  {\frac{{f_{d_{GF}^{}}}\left( y \right){f_{d_{GB}^{}}}\left( z \right)}{{\ln 2}{\left(1 + t\right)}}} } } }\notag \\
 &\times \exp \left( { - \frac{{{P_{GF}}\gamma _{GB}^{th}x{y^{ - \alpha }} + {\sigma ^2}\gamma _{GB}^{th}}}{{{\lambda _{GB}}{P_{GB}}{z^{ - \alpha }}}}} \right){f_{{{\left| {{h_{GF}}} \right|}^2}}}\left( x \right)dxdydzdt.
\end{align}}

{Based on Eq. [1.3.2.22] and Eq. [2.3.4.3] in \cite{Prudnikov}, two integrals of \eqref{C1} can be derived. Then, utilizing Chebyshev-Gauss quadratures in \eqref{C1}, the final expressions can be derived. By this way, the expressions of the ergodic rate for GF users have been proved.}


\vspace{-0.3cm}
\begin{thebibliography}{00}


\bibitem{st32}
{Z. Ding, R. Schober, P. Fan and H. V. Poor, ``Simple semi-grant-free transmission strategies assisted by non-orthogonal multiple access," \emph{IEEE Trans. Commun.}, vol. 67, no. 6, pp. 4464-4478, Jun. 2019.}

\bibitem{semigf2}
{Z. Ding, R. Schober, H. V. Poor, ``A new QoS-guarantee strategy for NOMA assisted semi-grant-free transmission," \emph{arXiv preprint arXiv:2004.12997}, 2020.}

\bibitem{GFGB}
{N. Ksairi and M. Debbah, ``Uplink pilots for multiuser MIMO with mixed grant free and grant based transmissions," \emph{Proc. IEEE Veh. Tech. Conf.}, Porto, pp. 1-6, 2018.}

\bibitem{yuanwei5GNOMA}
{Y. Liu, Z. Qin, M. Elkashlan, Z. Ding, A. Nallanathan and L. Hanzo, ``Non-orthogonal multiple access for 5G and beyond," \emph{Proc. {IEEE}}, vol. 105, no. 12, pp. 2347-2381, Dec. 2017.}

\bibitem{8635489}
{W. Yi, Y. Liu, A. Nallanathan and M. Elkashlan, ``Clustered millimeter-wave Networks with non-orthogonal multiple access," \emph{IEEE Trans. Commun.}, vol. 67, no. 6, pp. 4350-4364, Jun. 2019.}

\bibitem{ICSI2}
{M. S. Ali, H. Tabassum and E. Hossain, ``Dynamic user clustering and power allocation for uplink and downlink non-orthogonal multiple access (NOMA) systems," \emph{IEEE Access}, vol. 4, pp. 6325-6343, 2016.}


\bibitem{TWC}
{Y. Liu, Z. Qin, M. Elkashlan, Y. Gao and L. Hanzo, ``Enhancing the physical layer security of non-orthogonal multiple access in large-scale networks," \emph{IEEE Trans. Wireless Commun.}, vol. 16, no. 3, pp. 1656-1672, Mar. 2017.}


\bibitem{Zhiguo}
{Z. Ding, R. Schober and H. V. Poor, ``Unveiling the importance of SIC in NOMA systems: Part I - state of the art and recent findings," \emph{IEEE Commun. Lett.}, doi: 10.1109/LCOMM.2020.3012604.}


\bibitem{Prudnikov}
{A. P. Prudnikov, Yu. A. Brychkov, and O. I. Marichev, \emph{Integrals and series: special functions}, vol. 1. New York: Gordon \& Breach Sci. Publ., 1986.}
\end{thebibliography}
\end{document}